\pdfoutput=1
\documentclass{llncs}

\usepackage[T1]{fontenc}
\usepackage[utf8x]{inputenc}
\usepackage{amssymb,color,relsize}
\definecolor{keywordcolor}{rgb}{0.7, 0.1, 0.1}   
\definecolor{tacticcolor}{rgb}{0.1, 0.2, 0.6}    
\definecolor{commentcolor}{rgb}{0.4, 0.4, 0.4}   
\definecolor{symbolcolor}{rgb}{0.0, 0.1, 0.6}    
\definecolor{sortcolor}{rgb}{0.1, 0.5, 0.1}      
\definecolor{attributecolor}{rgb}{0.7, 0.1, 0.1} 

\usepackage{listings}

\lstMakeShortInline"

\usepackage{rotating}
\usepackage{tikz}
\usetikzlibrary{snakes,arrows,shapes}
\tikzset{
  every node/.style={rectangle,rounded corners=0.1cm}
}

\usepackage[hidelinks,bookmarksdepth=2]{hyperref}
\newcommand{\keywords}[1]{\par\addvspace\baselineskip
\noindent\keywordname\enspace\ignorespaces#1}

\begin{document}

\title{Homotopy Type Theory in Lean}

\author{Floris van Doorn\inst{1}
  \and Jakob von Raumer\inst{2}
  \and Ulrik Buchholtz\inst{3}}

\institute{%
  Carnegie Mellon University, Pittsburgh, USA\\
  \email{fpvdoorn@gmail.com}
  \and
  University of Nottingham, Nottingham, United Kingdom\\
  \email{jakob@von-raumer.de}
  \and
  TU Darmstadt, Darmstadt, Germany\\
  \email{ulrikbuchholtz@gmail.com}
}

\maketitle

\begin{abstract}
  We discuss the homotopy type theory library in the Lean proof
  assistant. The library is especially geared toward synthetic homotopy theory.
  Of particular interest is the use of just a few primitive notions of higher
    inductive types, namely quotients and truncations, and the use of cubical methods.
  \keywords{Homotopy Type Theory \textperiodcentered{}
   Formalized Mathematics \textperiodcentered{}
   Lean \textperiodcentered{}
   Proof Assistants}
\end{abstract}

\section{Introduction}
\label{sec:introduction}

Homotopy type theory (HoTT) refers to the homotopical interpretation of
Martin-L{\"o}f's dependent type theory \cite{AwodeyWarren2009,Voevodsky2006}, which
grew out of the groupoid model of \cite{HofmannStreicher1998}. In the
standard interpretation, every type-theoretical construct corresponds to a
homotopy-invariant construction on spaces. An important example is the identity
type, which corresponds to the path space construction.

Just like extensional type theory can be interpreted in a variety of
categories, for instance elementary toposes, it is expected that
homotopy type theory has homotopy-coherent interpretations in higher
toposes. Conversely, the interpretation has inspired new
type-theoretic ideas such as higher inductive types (HITs) and
Voevodsky's univalence axiom.
(See the HoTT book~\cite{hottbook} for more about HoTT.)

Most previous formalizations of HoTT used proof assistants that were not
originally designed with the homotopy interpretation in mind. In Coq we
have both Voevodsky et~al.'s \emph{UniMath} project \cite{UniMath} and the
HoTT library \cite{bauer2016coqhott}. In Agda, there is another
substantial HoTT library \cite{HoTTAgda}. The former library eschews
the use of HITs by instead using Voevodsky's resizing
axiom. Common for all of these libraries is that certain tricks
are used to accommodate HoTT: resizing is implemented bluntly in
UniMath using the inconsistent principle \emph{type-in-type}, while
HITs are implemented in the other libraries using ``Licata's trick''~\cite{Licata2011trick}. There is
also an impressive experimental proof assistant implementing cubical type theory~\cite{cubicaltt}
which is designed with the homotopy interpretation in mind, but it lacks many features that make a
proof assistant convenient to use, and the library is so far rudimentary.

\subsubsection{Contributions}
In this paper, we report on a new library%
\footnote{Available as part of: \url{https://github.com/leanprover/lean2}}
for HoTT in the proof assistant Lean~\cite{Moura2015}.
Lean is open source and implements dependent type theory.
It is designed to have a small kernel, with many features built outside the kernel.
We describe Lean in greater detail in Sect.~\ref{sec:lean}.
The "cloc" tool\footnote{\url{https://github.com/AlDanial/cloc}} reports the
library as having 30\,400 lines of specification and proof and 3\,600
lines of comments. Thus, our library is roughly the same size as the
Coq HoTT library, which has 29\,800 lines of specification and proof.
Our library includes many theorems from synthetic homotopy theory and a large algebraic hierarchy.
We describe the library in more detail in Sect.~\ref{sec:library}.
In the library we heavily use cubical techniques for higher path
algebra, see Sect.~\ref{sec:cubical}. We also have a novel approach
to implement HITs, which amounts to having two simple built-in HITs and
reducing everything else to those, as described in Sect.~\ref{sec:hits}.

\section{The Lean Proof Assistant}
\label{sec:lean}


Lean~\cite{Moura2015} is an interactive theorem prover which is mainly developed at Microsoft Research
and Carnegie Mellon University.
The project was started in 2013 by Leonardo de Moura
and has since gained the attention of academics as well as hands-on users.
Lean is an open-source program released under the Apache License 2.0 and welcomes
additions to its code and mathematical libraries.

In its short history, Lean has undergone several major changes.
The second version (Lean~2) supports two kernel modes.
The standard mode is for proof irrelevant reasoning, in which "Prop", the bottom universe,
contains types whose objects are considered to be judgmentally equal.
Since this is incompatible with homotopy type theory, a second HoTT mode was added, where
proof irrelevance is not present.
In 2016, the third major version of Lean (Lean~3) was released \cite{moura2017poplslides}.
In this version, many components of Lean have been rewritten. Of note, the unification procedure
has been restricted, since the full higher-order unification which is available in Lean~2 can lead to
timeouts and error messages that are unrelated to the actual mistakes.
Due to certain design decisions, such as proof erasure in
the virtual machine and a function definition package which requires axiom K~\cite{goguen2006eliminating},
the homotopy type theory mode is currently not supported in Lean~3.
This has led to the situation that the homotopy type theory library is kept in
the still maintained but not further developed Lean~2.
In the future we hope that we will find a way to support a version of homotopy
type theory in Lean~3 or a fork thereof.

The HoTT kernel of Lean~2 provides the following primitive notions:
\begin{itemize}
\item \emph{Type universes} "Type.{u} : Type.{u + 1}" for each universe level $u \in \mathbb{N}$.
  In Lean, this chain of universes is non-cumulative, and all universes are predicative.
  \clearpage
\item \emph{Function types} "A → B : Type.{max u v}" for types "A : Type.{u}" and "B : Type.{v}" as well as
\emph{dependent function types} "Πa, B a : Type.{max u v}" for each type "A : Type.{u}" and type family
"B : A → Type.{v}". These come with the usual "β" and "η" rules.
\item \emph{inductive types} and \emph{inductive type families}, as proposed by
Peter Dybjer~\cite{dybjer1994inductive}.
Every inductive definition adds its constructors and dependent recursors to the environment.
Pattern matching is \emph{not} part of the kernel
\item two kinds of \emph{higher inductive types}:
"n"-truncation and (typal) quotients (cf. Sect.~\ref{sec:hits}).
\end{itemize}

Outside the kernel, Lean's elaborator uses backtracking search to infer implicit information. It
does the following simultaneously.
\begin{itemize}
\item The elaborator fills in \emph{implicit arguments}, which can be inferred from the context,
  such as the type of the term to be constructed and the given explicit arguments. Users mark
  implicit arguments with curly braces. For example, the type of equality is
  "eq : Π{A : Type}, A → A → Type", which allows the user to write "eq a₁ a₂" or "a₁ = a₂" instead
  of "@eq A a₁ a₂". The symbol "@" allows the user to fill in implicit arguments explicitly. The
  elaborator supports both first-order unification and higher-order unification.
\item We can mark functions as \emph{coercions}, which are then ``silently'' applied when needed.
  For example, we have equivalences "f : A ≃ B", which is a structure consisting of a function
  "A → B" with a proof that the function is an equivalence. The map "(A ≃ B) → (A → B)" is marked as
  a coercion. This means that we can write "f a" for "f : A ≃ B" and "a : A", and the coercion is
  inserted automatically.
\item Lean was designed with \emph{type classes} in mind, which can provide canonical inhabitants of
  certain types.  This is especially useful for algebraic structures (see
  Sect.~\ref{sec:library-algebra}) and for type properties like truncatedness and connectedness.
  Type class instances can refer to other type classes, so that we can chain them together. This
  makes it possible for Lean to automatically infer why types are "n"-truncated if our reasoning
  requires this, for example when we are eliminating out of a truncated type. For example we show
  that the type of functors between categories "C" and "D" is equivalent to an iterated sigma type. 
  \begin{lstlisting}[gobble=2]
  (Σ (F₀ : C → D) (F₁ : Π {a b}, hom a b → hom (F₀ a) (F₀ b)),
  (Π (a), F₁ (ID a) = ID (F₀ a)) ×
  (Π {a b c} (g : hom b c) (f : hom a b),
    F₁ (g ∘ f) = F₁ g ∘ F₁ f)) ≃ functor C D
  \end{lstlisting}
  Note the use of coercions here: "F₀ : C → D" really means a function from the objects of "C" to
  the objects of "D". From this equivalence, Lean's type class inference can automatically infer
  that "functor C D" is a set if the objects of "D" form a set. Type class inference will repeatedly
  apply the rules when sigma-types and pi-types are sets, and use the facts that hom-sets are sets
  and that equalities in sets are sets (in total 20 rules are applied for this example).
\item Instead of giving constructions by explicit terms, we can also make use of
Lean's \emph{tactics}, which give us an alternative way to construct terms step by
step. This is especially useful if the proof term is large, or if the elaboration relies heavily on
higher-order unification.
\item We can define custom syntax, including syntax with binding.
In the following example we declare two custom notations.
\begin{lstlisting}
infix ⬝ := concat
notation `Σ` binders `, ` r:(scoped P, sigma P) := r
\end{lstlisting}
The first line allows us to write "p ⬝ q" for path concatenation "concat p q". The second line
allows us to write "Σ x, P x" instead of "sigma P". This notation can also be chained:
"Σ (A : Type) (a : A), a = a" means "sigma (λ(A : Type), sigma (λ(a : A), a = a))".
\end{itemize}

\subsection{Consistency of HoTT Lean}
\label{sec:consistency}

Voevodsky's model of univalence in simplicial sets \cite{Kapulkin-Lumsdaine2012} covers the type
theory with empty, unit, disjoint sums, pi, sigma, identity, and W-types and one univalent universe
\`a la Tarski closed under the these type formers.
The model validates the "β" and "η" rules for function types.

The cubical type theory of \cite{Cohen-Coquand-Huber-Mortberg2016} interprets Martin-L{\"o}f type
theory using Andrew Swan's construction of the identity types.
(The cubical path types of this model do not satisfy the computation rule for identity types.)
It has been checked that the corresponding model in cubical sets based on de~Morgan algebras models two
HITs, namely suspension and propositional truncation.
(The model even satisfies the computation rules
for the path constructors.)
The technique used also covers pushouts, so by the reduction of
"n"-truncation to pushouts \cite{rijke2017join}, the models covers all "n"-truncations.  We believe
this model also covers all the ordinary inductive families supported by Lean, but this has not been
checked in detail.

Mark Bickford's formalization of the cubical
model\footnote{\url{http://www.nuprl.org/wip/Mathematics/cubical!type!theory/}} covers a whole
hierarchy of universes like we have in the Lean kernel.  It additionally verifies some novel type
constructors such as a higher dimensional intersection type.

These models provide us with high confidence that the logic implemented by the Lean HoTT kernel is
consistent. Furthermore, the kernel is very small compared to other kernels implementing dependent
type theory. The kernel does not contain pattern matching, a termination checker, fixpoint
operations or module management. This increases the confidence that the kernel implements the logic correctly.
Furthermore, the only thing we do outside the kernel to extend the logic is to posit the univalence
axiom; we do not use type-in-type or Licata's trick or anything else which might introduce inconsistencies.

\section{The Structure of the Library}
\label{sec:library}

In this section we describe the overall structure of the homotopy type theory library and we
highlight some examples.

The library contains a markdown file in each folder to describe the contents of the files in that
folder.  For readers familiar with~\cite{hottbook}, the library includes a
file\footnote{\url{https://github.com/leanprover/lean2/blob/master/hott/book.md}} "book.md" that
describes where in the library the various parts of the book are formalized.

Figures \ref{fig:init}-\ref{fig:category} contain graphs of the files in various parts of the
library; the edges denote the dependencies of the files. Each folder contains a file "default"
which only contains imports of various files in the folder and which is imported if the user imports
the folder.
There are also three additional folders: "types" (see Subsect.~\ref{sec:library-types}), "cubical",
related to the cubical methods discussed in Sect.~\ref{sec:cubical}; and "hit", related to higher
inductive types as discussed in Sect.~\ref{sec:hits}.
There are also some files in the root folder which we do not describe here.

There is a separate "Spectral" repository,\footnote{It has 7\,700 lines of code and 1\,400 lines of
  comments. It is available at \url{https://github.com/cmu-phil/Spectral}} the goal of which is to
formalize the Serre Spectral Sequence, and which will be merged into the Lean-HoTT library in the
future. Some examples below are located in this repository.

\begin{figure}
\centering
\begin{minipage}[b]{.5\textwidth}
\centering
\begin{tikzpicture}[scale=0.7,>=latex,line join=bevel,]
\node (hedberg) at (128.0bp,580.0bp) [draw] {hedberg};
  \node (num) at (169.0bp,182.0bp) [draw] {num};
  \node (relation) at (104.0bp,182.0bp) [draw] {relation};
  \node (equiv) at (229.0bp,410.0bp) [draw] {equiv};
  \node (bool) at (166.0bp,125.0bp) [draw] {bool};
  \node (nat) at (169.0bp,410.0bp) [draw] {nat};
  \node (pointed) at (225.0bp,638.0bp) [draw] {pointed};
  \node (function) at (109.0bp,295.0bp) [draw] {function};
  \node (wf) at (324.0bp,638.0bp) [draw] {wf};
  \node (hit) at (165.0bp,638.0bp) [draw] {hit};
  \node (tactic) at (169.0bp,238.5bp) [draw] {tactic};
  \node (pathover) at (229.0bp,468.0bp) [draw] {pathover};
  \node (util) at (268.0bp,580.0bp) [draw] {util};
  \node (path) at (169.0bp,352.0bp) [draw] {path};
  \node (datatypes) at (136.0bp,11.0bp) [draw] {datatypes};
  \node (types) at (106.0bp,238.5bp) [draw] {types};
  \node (trunc) at (224.0bp,524.0bp) [draw] {trunc};
  \node (default) at (195.0bp,695.0bp) [draw] {default};
  \node (reserved_notation) at (136.0bp,68.0bp) [draw] {reserved\ notation};
  \node (logic) at (109.0bp,125.0bp) [draw] {logic};
  \node (connectives) at (40.0bp,580.0bp) [draw] {connectives};
  \node (ua) at (311.0bp,468.0bp) [draw] {ua};
  \node (funext) at (326.0bp,580.0bp) [draw] {funext};
  \draw [->] (hedberg) ..controls (160.26bp,561.18bp) and (184.73bp,546.91bp)  .. (trunc);
  \draw [->] (pointed) ..controls (224.73bp,607.2bp) and (224.38bp,567.26bp)  .. (trunc);
  \draw [->] (reserved_notation) ..controls (136.0bp,50.785bp) and (136.0bp,41.186bp)  .. (datatypes);
  \draw [->] (default) ..controls (204.16bp,677.59bp) and (209.69bp,667.09bp)  .. (pointed);
  \draw [->] (default) ..controls (168.75bp,676.81bp) and (147.81bp,662.12bp)  .. (130.0bp,649.0bp) .. controls (105.62bp,631.04bp) and (78.04bp,609.75bp)  .. (connectives);
  \draw [->] (pathover) ..controls (229.0bp,449.6bp) and (229.0bp,439.97bp)  .. (equiv);
  \draw [->] (logic) ..controls (117.63bp,106.78bp) and (122.48bp,96.549bp)  .. (reserved_notation);
  \draw [->] (hit) ..controls (179.93bp,609.16bp) and (202.05bp,566.41bp)  .. (trunc);
  \draw [->] (default) ..controls (185.64bp,677.21bp) and (179.74bp,666.02bp)  .. (hit);
  \draw [->] (ua) ..controls (290.27bp,453.34bp) and (267.16bp,436.99bp)  .. (equiv);
  \draw [->] (nat) ..controls (169.0bp,393.53bp) and (169.0bp,382.76bp)  .. (path);
  \draw [->] (path) ..controls (169.0bp,321.44bp) and (169.0bp,282.0bp)  .. (tactic);
  \draw [->] (tactic) ..controls (169.0bp,221.13bp) and (169.0bp,209.92bp)  .. (num);
  \draw [->] (path) ..controls (150.29bp,334.22bp) and (137.21bp,321.8bp)  .. (function);
  \draw [->] (funext) ..controls (293.81bp,562.32bp) and (265.63bp,546.86bp)  .. (trunc);
  \draw [->] (types) ..controls (105.37bp,220.68bp) and (105.02bp,210.95bp)  .. (relation);
  \draw [->] (relation) ..controls (105.51bp,164.78bp) and (106.35bp,155.19bp)  .. (logic);
  \draw [->] (default) ..controls (235.86bp,676.94bp) and (278.67bp,658.03bp)  .. (wf);
  \draw [->] (default) ..controls (168.86bp,678.06bp) and (152.27bp,664.8bp)  .. (143.0bp,649.0bp) .. controls (134.49bp,634.48bp) and (130.79bp,615.69bp)  .. (hedberg);
  \draw [->] (trunc) ..controls (207.77bp,506.81bp) and (196.1bp,492.93bp)  .. (189.0bp,479.0bp) .. controls (180.95bp,463.22bp) and (175.68bp,443.71bp)  .. (nat);
  \draw [->] (function) ..controls (108.09bp,277.85bp) and (107.58bp,268.18bp)  .. (types);
  \draw [->] (default) ..controls (229.03bp,680.6bp) and (251.05bp,667.77bp)  .. (262.0bp,649.0bp) .. controls (270.59bp,634.27bp) and (271.5bp,614.68bp)  .. (util);
  \draw [->] (wf) ..controls (324.61bp,620.34bp) and (324.97bp,609.75bp)  .. (funext);
  \draw [->] (bool) ..controls (156.89bp,107.69bp) and (151.15bp,96.789bp)  .. (reserved_notation);
  \draw [->] (trunc) ..controls (225.42bp,508.06bp) and (226.31bp,498.16bp)  .. (pathover);
  \draw [->] (util) ..controls (254.54bp,562.87bp) and (244.97bp,550.69bp)  .. (trunc);
  \draw [->] (connectives) ..controls (42.938bp,551.09bp) and (47.0bp,505.99bp)  .. (47.0bp,468.0bp) .. controls (47.0bp,468.0bp) and (47.0bp,468.0bp)  .. (47.0bp,352.0bp) .. controls (47.0bp,314.52bp) and (72.637bp,277.14bp)  .. (types);
  \draw [->] (funext) ..controls (322.07bp,550.68bp) and (316.63bp,510.02bp)  .. (ua);
  \draw [->] (equiv) ..controls (209.98bp,391.61bp) and (196.97bp,379.04bp)  .. (path);
  \draw [->] (types) ..controls (126.41bp,220.19bp) and (141.69bp,206.49bp)  .. (num);
  \draw [->] (num) ..controls (168.17bp,166.21bp) and (167.59bp,155.17bp)  .. (bool);
\end{tikzpicture}
  \caption{The initial part of the library}
  \label{fig:init}
\end{minipage}%
\begin{minipage}[b]{.5\textwidth}
  \centering
\begin{tikzpicture}[scale=0.7,>=latex,line join=bevel]
\node (binary) at (38.0bp,67.0bp) [draw] {binary};
  \node (ordered_ring) at (42.0bp,296.0bp) [draw] {ordered\ ring};
  \node (ordered_group) at (38.0bp,238.0bp) [draw] {ordered\ group};
  \node (inf_group) at (130.0bp,125.0bp) [draw] {inf.\ group};
  \node (priority) at (129.0bp,67.0bp) [draw] {priority};
  \node (field) at (141.0bp,296.0bp) [draw] {field};
  \node (group) at (130.0bp,181.5bp) [draw] {group};
  \node (ring) at (141.0bp,238.0bp) [draw] {ring};
  \node (order) at (38.0bp,125.0bp) [draw] {order};
  \node (ordered_field) at (91.0bp,353.0bp) [draw] {ordered\ field};
  \draw [->] (ordered_ring) -- (ordered_group);
  \draw [->] (inf_group) -- (binary);
  \draw [->] (ordered_group) -- (group);
  \draw [->] (order) -- (binary);
  \draw [->] (ordered_field) -- (ordered_ring);
  \draw [->] (ordered_field) -- (field);
  \draw [->] (ordered_ring) -- (ring);
  \draw [->] (group) -- (inf_group);
  \draw [->] (ring) -- (group);
  \draw [->] (inf_group) -- (priority);
  \draw [->] (field) -- (ring);
  \draw [->] (ordered_group) -- (order);
  \draw [->] (order) -- (priority);
\end{tikzpicture}
  \caption{The algebraic hierarchy}
  \label{fig:algebra}
\end{minipage}
\end{figure}

\subsection{The initial part of the library}
\label{sec:library-init}

Figure~\ref{fig:init} illustrates the files of the initial part of the library. These files are
imported by default when opening a Lean file.
The very first file, "datatypes", defines the basic datatypes, such as
"unit", "empty", "eq", "prod", "sum", "sigma", "bool", "nat". Higher up, the "path" file develops the basic
properties of the identity type (also called equality or identification type) in HoTT.
This includes the basic properties of homotopies, transport and the low-dimensional "∞"-groupoid structure of types.

In the rest of the files we define equivalences, posit the univalence axiom and derive function
extensionality from univalence (in "equiv", "ua" and "funext", respectively).
However, in order to be able to track which definitions only depend on function extensionality and
not univalence, via the "print axioms" command, we also add function extensionality directly as an axiom.

Lastly, we develop "n"-truncated types, initialize the primitive HITs,
prove that types with decidable equality are sets~\cite{Hedberg1998} and define
the basic notions of pointed types (in "trunc", "hit", "hedberg" and "pointed", respectively).

\subsection{Facts about Types}
\label{sec:library-types}

The files in subdirectory "types" develop in more detail the properties and constructions related to
individual types and type formers. For types like "sum", "sigma" and "pi" we characterize the
equality in that type, define the functorial action and show that the functorial action preserves equivalences.
In "univ" we prove properties of type universes, such as the object classifier property.
Of particular importance is the file "pointed", which contains properties of pointed
types, maps, equivalences and homotopies, which contains over 2\,000 lines (also counting the
corresponding file in the Spectral repository).

\subsection{The Algebraic Hierarchy}
\label{sec:library-algebra}

The algebraic hierarchy, all in the "algebra" subdirectory, is structured as seen
in figure~\ref{fig:algebra}.  That figure does not contain files that depend on the category theory
sublibrary. The algebraic hierarchy defines common algebraic structures, starting with small
structures, like semigroups and partial orders, and extending them to groups, rings, all the
way up to discrete linear ordered fields.
(Discrete means that the order is decidable.)

We combine the ``partially bundled'' approach with the ``fully bundled'' approach in the algebraic
hierarchy, similar to how algebraic structres in the Coq library are defined~\cite{bundled}.
The partially bundled approach means that given a type "A" we define what it means that
"A" has a group-structure or ring-structure. This is used for concrete structures, and we use type
classes to infer these inhabitants. For example, we prove that "ℕ" forms a decidable linear ordered
semiring, and mark this as an instance. If we want to show that for "n m k : ℕ" we have
"(n * m) * k = n * (m * k)", we can use "mul.assoc", the theorem that multiplication in any
semigroup is associative. Then type class inference will try to show that "ℕ" is a semigroup, and it
will use the instance that every decidable linear ordered semiring is a semigroup. We use Lean's "extend"
syntax to easily define new algebraic structures. For example, the following code defines a
structure "ab_group" of abelian groups, which consists of the fields of both "group" and
"comm_monoid". Also, the instances "ab_group A → group A" and "ab_group A → comm_monoid A" are
automatically generated.
\begin{lstlisting}
structure ab_group [class] (A : Type)
  extends group A, comm_monoid A
\end{lstlisting}
We use the fully bundled approach when doing group theory and other algebra. A bundled structure is
a type together with a structure on that type. For example, this is the definition of a bundled group:
\begin{lstlisting}
structure Group := (carrier : Type) (struct : group carrier)
\end{lstlisting}
We define "Group.carrier" to be a coercion. We make "Group.struct" an instance, which means that if we have to synthesize a term of type "group (Group.carrier G)", Lean will automatically find this instance. We use the bundled structures for group theory. For example, if "G H : Group" then we define the product group "G ×g H". We use "×g"
for the product of two groups to disambiguate it from other products, like the product of two types,
two pointed types or two truncated types (type class inference does not work well to disambiguate here,
since all these structures coerce to types).

If we go back to the example "(n * m) * k = n * (m * k)" on "ℕ", we also interpret the
multiplication symbol on "ℕ" using type class inference. In this case, Lean will try to find an
instance of "has_mul ℕ", where "has_mul" is a type class stating that the type has a multiplication.
Lean can find this instance since we have a general instance "semigroup A → has_mul A". However,
since we want to also have additive semigroups, we have a different notion of additive semigroups,
"add_semigroup", with corresponding instance "add_semigroup A → has_add A". To minimize overhead, we
can define additive structures as the multiplicative counterpart, and then prove theorems about
additive structures by using the corresponding theorem for multiplicative structures. We do have to
manually define the instances for additive structures. Here is an example for semigroups:
\begin{lstlisting}
definition add_semigroup [class] : Type → Type := semigroup
definition has_add_of_add_semigroup [instance] (A : Type)
  [s : add_semigroup A] : has_add A :=
has_add.mk (@semigroup.mul A s)
definition add.assoc {A : Type} [s : add_semigroup A]
  (a b c : A) : (a + b) + c = a + (b + c) :=
@mul.assoc A s a b c
\end{lstlisting}
This approach has advantages and disadvantages. An advantage is that theorem names are different for
additive structures and multiplicative structures, so we can write "add.assoc" for associativity of
addition and "mul.assoc" for associativity of multiplication. Furthermore, we can easily define a
ring by extending an additive abelian group and a multiplicative monoid (plus distributivity).

A disadvantage is that operations that are traditionally not written using "+" or "*", such as
concatenating two lists, do not fall in either category.
Also, in our formalization we make a 
distinction between additive and multiplicative groups. Since we define additive groups as
multiplicative groups, we can still apply theorems about multiplicative groups to additive groups,
but some care is needed when doing this: if one applies a theorem about multiplicative groups with
assumption "n * k = 1" to an additive group, the new subgoal becomes "n * k = 1", even though in an
additive group this really means "n + k = 0".

All the algebraic structures we mentioned so far (not including "has_mul" and "has_add") are assumed
to be sets, i.e., 0-truncated. We also have variants of some of these structures which are not assumed to
be sets. For example, we have "inf_group" and "inf_ab_group", which are (abelian) groups without the
assumption that they are sets, but without higher coherences.
This is useful for, e.g., loop spaces or pointed maps into loop
spaces, since those types are not groups, but will become groups (the homotopy and cohomology
groups) after applying set-truncation.


\subsection{Homotopy Theory}
\label{sec:library-homotopy}

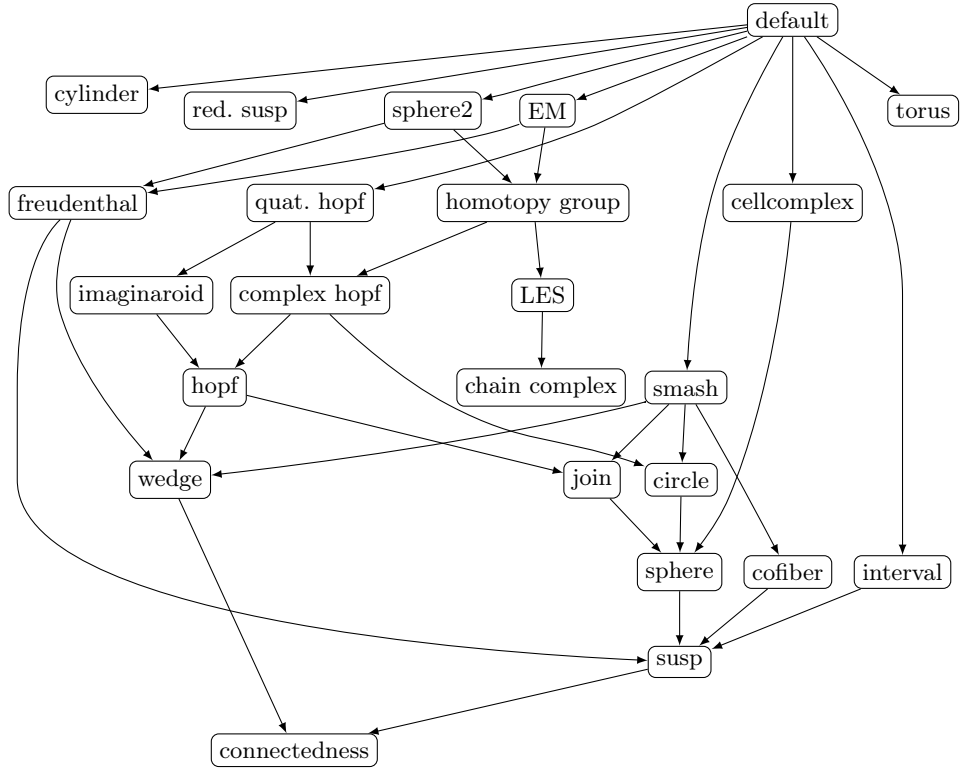
\begin{figure}[t]
  \centering
\begin{tikzpicture}[scale=0.6,>=latex,line join=bevel,]

\node (wedge) at (126.0bp,180.0bp) [draw] {wedge};
  \node (cylinder) at (80.0bp,422.0bp) [draw] {cylinder};
  \node (cellcomplex) at (517.0bp,354.0bp) [draw] {cellcomplex};
  \node (sphere) at (446.0bp,122.0bp) [draw] {sphere};
  \node (chain_complex) at (359.0bp,238.0bp) [draw] {chain complex};
  \node (complex_hopf) at (214.0bp,296.0bp) [draw] {complex hopf};
  \node (sphere2) at (291.0bp,412.0bp) [draw] {sphere2};
  \node (connectedness) at (204.0bp,10.0bp) [draw] {connectedness};
  \node (circle) at (447.0bp,180.0bp) [draw] {circle};
  \node (hopf) at (154.0bp,238.0bp) [draw] {hopf};
  \node (EM) at (363.0bp,412.0bp) [draw] {EM};
  \node (cofiber) at (514.0bp,122.0bp) [draw] {cofiber};
  \node (freudenthal) at (68.0bp,354.0bp) [draw] {freudenthal};
  \node (red_susp) at (170.0bp,412.0bp) [draw] {red. susp};
  \node (quaternionic_hopf) at (214.0bp,354.0bp) [draw] {quat.\ hopf};
  \node (imaginaroid) at (108.0bp,296.0bp) [draw] {imaginaroid};
  \node (LES_of_homotopy_groups) at (360.0bp,296.0bp) [draw] {LES};
  \node (smash) at (450.0bp,238.0bp) [draw] {smash};
  \node (join) at (391.0bp,180.0bp) [draw] {join};
  \node (susp) at (446.0bp,65.5bp) [draw] {susp};
  \node (default) at (517.0bp,469.0bp) [draw] {default};
  \node (torus) at (599.0bp,412.0bp) [draw] {torus};
  \node (interval) at (586.0bp,122.0bp) [draw] {interval};
  \node (homotopy_group) at (354.0bp,354.0bp) [draw] {homotopy group};
  \draw [->] (complex_hopf) ..controls (240.45bp,270.12bp) and (270.65bp,243.02bp)  .. (301.0bp,227.0bp) .. controls (348.74bp,201.8bp) and (365.87bp,208.3bp)  .. (417.0bp,191.0bp) .. controls (418.21bp,190.59bp) and (419.45bp,190.16bp)  .. (circle);
  \draw [->] (imaginaroid) ..controls (123.11bp,276.95bp) and (132.08bp,265.64bp)  .. (hopf);
  \draw [->] (EM) ..controls (360.38bp,395.11bp) and (358.77bp,384.76bp)  .. (homotopy_group);
  \draw [->] (smash) ..controls (449.09bp,220.34bp) and (448.54bp,209.75bp)  .. (circle);
  \draw [->] (smash) ..controls (431.61bp,219.92bp) and (418.08bp,206.62bp)  .. (join);
  \draw [->] (homotopy_group) ..controls (355.9bp,335.6bp) and (356.9bp,325.97bp)  .. (LES_of_homotopy_groups);
  \draw [->] (hopf) ..controls (145.2bp,219.77bp) and (140.16bp,209.33bp)  .. (wedge);
  \draw [->] (cellcomplex) ..controls (513.15bp,313.17bp) and (503.24bp,231.4bp)  .. (477.0bp,169.0bp) .. controls (472.83bp,159.07bp) and (466.51bp,148.99bp)  .. (sphere);
  \draw [->] (sphere2) -- (freudenthal);
  \draw [->] (hopf) ..controls (212.77bp,223.62bp) and (316.94bp,198.12bp)  .. (join);
  \draw [->] (smash) ..controls (426.92bp,229.73bp) and (421.81bp,228.18bp)  .. (417.0bp,227.0bp) .. controls (324.98bp,204.39bp) and (213.84bp,189.9bp)  .. (wedge);
  \draw [->] (freudenthal) ..controls (38.25bp,325.66bp) and (30.0bp,282.82bp)  .. (30.0bp,238.0bp) .. controls (30.0bp,238.0bp) and (30.0bp,238.0bp)  .. (30.0bp,180.0bp) .. controls (30.0bp,93.508bp) and (318.66bp,71.426bp)  .. (susp);
  \draw [->] (homotopy_group) ..controls (306.92bp,334.49bp) and (272.04bp,320.05bp)  .. (complex_hopf);
  \draw [->] (smash) ..controls (461.41bp,218.48bp) and (469.87bp,203.83bp)  .. (477.0bp,191.0bp) .. controls (486.34bp,174.17bp) and (496.67bp,154.84bp)  .. (cofiber);
  \draw [->] (default) -- (cylinder);
  \draw [->] (default) ..controls (473.15bp,444.4bp) and (410.86bp,409.98bp)  .. (388.0bp,401.0bp) .. controls (347.31bp,385.01bp) and (299.21bp,372.47bp)  .. (quaternionic_hopf);
  \draw [->] (default) ..controls (500.59bp,441.02bp) and (478.86bp,401.33bp)  .. (468.0bp,365.0bp) .. controls (457.05bp,328.37bp) and (452.69bp,283.74bp)  .. (smash);
  \draw [->] (default) ..controls (541.75bp,433.35bp) and (586.0bp,362.29bp)  .. (586.0bp,296.0bp) .. controls (586.0bp,296.0bp) and (586.0bp,296.0bp)  .. (586.0bp,238.0bp) .. controls (586.0bp,204.58bp) and (586.0bp,165.74bp)  .. (interval);
  \draw [->] (default) -- (EM);
  \draw [->] (default) -- (torus);
  \draw [->] (sphere2) -- (homotopy_group);
  \draw [->] (default) -- (cellcomplex);
  \draw [->] (interval) ..controls (542.37bp,104.39bp) and (497.77bp,86.392bp)  .. (susp);
  \draw [->] (LES_of_homotopy_groups) ..controls (359.68bp,277.6bp) and (359.52bp,267.97bp)  .. (chain_complex);
  \draw [->] (quaternionic_hopf) ..controls (214.0bp,335.6bp) and (214.0bp,325.97bp)  .. (complex_hopf);
  \draw [->] (quaternionic_hopf) ..controls (177.89bp,334.24bp) and (153.7bp,321.0bp)  .. (imaginaroid);
  \draw [->] (join) ..controls (407.77bp,162.31bp) and (419.64bp,149.8bp)  .. (sphere);
  \draw [->] (freudenthal) ..controls (58.375bp,330.4bp) and (50.881bp,305.45bp)  .. (57.0bp,285.0bp) .. controls (67.23bp,250.81bp) and (92.571bp,217.43bp)  .. (wedge);
  \draw [->] (susp) ..controls (391.38bp,52.974bp) and (300.29bp,32.084bp)  .. (connectedness);
  \draw [->] (circle) ..controls (446.7bp,162.61bp) and (446.53bp,152.49bp)  .. (sphere);
  \draw [->] (cofiber) ..controls (492.46bp,104.1bp) and (476.16bp,90.561bp)  .. (susp);
  \draw [->] (wedge) ..controls (143.91bp,140.96bp) and (178.53bp,65.505bp)  .. (connectedness);
  \draw [->] (EM) ..controls (346.58bp,404.04bp) and (342.18bp,402.26bp)  .. (338.0bp,401.0bp) .. controls (296.6bp,388.49bp) and (177.87bp,370.05bp)  .. (freudenthal);
  \draw [->] (default) -- (red_susp);
  \draw [->] (complex_hopf) ..controls (193.93bp,276.6bp) and (181.58bp,264.66bp)  .. (hopf);
  \draw [->] (default) ..controls (461.01bp,455.26bp) and (394.28bp,438.59bp)  .. (338.0bp,423.0bp) .. controls (336.36bp,422.54bp) and (334.68bp,422.07bp)  .. (sphere2);
  \draw [->] (sphere) ..controls (446.0bp,103.59bp) and (446.0bp,93.946bp)  .. (susp);
\end{tikzpicture}

\caption{The homotopy theory part of the library}
\label{fig:homotopy}
\end{figure}

The homotopy theory part of the library is organized as shown in figure~\ref{fig:homotopy}. Almost
all results in Chapter 8 of the HoTT book have been formalized in Lean. In particular it contains
various results about connectedness, a version of the Freudenthal suspension theorem, the complex
and quaternionic Hopf fibration~\cite{buchholtz-rijke-2016} and the long exact sequence of homotopy
groups. Together these results show:
\begin{lstlisting}
definition πnSn (n : ℕ) : πg[n+1] (S* (n+1)) ≃g gℤ
definition π3S2 : πg[3] (S* 2) ≃g gℤ
\end{lstlisting}
This is to say that the "n"-th homotopy group of the "n"-sphere (for "n ≥ 1") and the 3\textsuperscript{rd} homotopy group of the
2-sphere are group isomorphic to the integers. Of note here is the notation "πg[n] A"
which denotes the "n"-th homotopy group of "A", as a group. In contrast, we also have the operation
"π[n] A" which is the "n"-th homotopy group of "A" as a pointed type, which is also
defined for "n = 0".  Originally, we defined "ghomotopy_group : ℕ → Type → Group" where
"ghomotopy_group n A" is the "(n+1)"-st homotopy group of "A" and we had notation "πg[n+1] A" for
this. However, this requires the user to write the third homotopy group as "πg[2+1]". To remedy
this, we changed the definition of "ghomotopy_group" to have type
"Π(n : ℕ) [H : is_succ n], Type → Group", where "H" is a proof that "n" is a successor
of a natural number, and which is synthesized using type class inference.

We also prove Whitehead's principle for truncated types and the Seifert-van Kampen theorem, and we
define the Eilenberg-Maclane spaces and show that they are unique. Furthermore, we define operations
on types of homotopy theoretic significance, such as cofibers, joins, and wedge and smash products,
and prove various properties about them, such as the associativity of the join and smash products
and the fact that the suspension and smash product have right adjoints, respectively loop spaces and
pointed maps.

\subsection{Category Theory}
\label{sec:library-category}

It seems a constant across many libraries of formalized mathematics that the development of category
theory takes up a substantial fraction of the files, and our library is the same way, as can be seen
in figure~\ref{fig:category}. Highlights include the Yoneda lemma and the Rezk completion.~\cite{ahrens2015rezk}

As an example from this part of the library, consider this excerpt
which formalizes the fact that the Yoneda embedding preserves existing limits:
\begin{lstlisting}[gobble=2]
  definition yoneda_embedding (C : Precategory)
    : C ⇒ cset ^c Cᵒᵖ

  variables {C D : Precategory}
  definition preserves_existing_limits [class] (G : C ⇒ D) :=
  Π(I : Precategory) (F : I ⇒ C)
    [H : has_terminal_object (cone F)],
    is_terminal (cone_obj_compose G (terminal_object (cone F)))

  theorem preserves_existing_limits_yoneda_embedding
    (C : Precategory)
    : preserves_existing_limits (yoneda_embedding C)
\end{lstlisting}

\begin{sidewaysfigure}
  \centering
\begin{tikzpicture}[scale=0.7,>=latex,line join=bevel,]
\node (c*opposite) at (405.95bp,236.0bp) [draw] {c.opposite};
  \node (f*attributes) at (642.95bp,178.5bp) [draw] {f.attributes};
  \node (l*adjoint) at (556.95bp,468.0bp) [draw] {l.adjoint};
  \node (groupoid) at (218.95bp,236.0bp) [draw] {groupoid};
  \node (l*functor_preserve) at (744.95bp,468.0bp) [draw] {l.functor\ preserve};
  \node (c*product) at (665.95bp,294.0bp) [draw] {c.product};
  \node (l*colimits) at (637.95bp,410.0bp) [draw] {l.colimits};
  \node (c*finite_cats) at (740.95bp,178.5bp) [draw] {c.finite\ cats};
  \node (c*fundamental_groupoid) at (92.951bp,294.0bp) [draw] {c.fund.\ groupoid};
  \node (f*yoneda) at (738.95bp,410.0bp) [draw] {f.yoneda};
  \node (category) at (494.95bp,178.5bp) [draw] {category};
  \node (f*examples) at (735.95bp,352.0bp) [draw] {f.examples};
  \node (strict) at (564.95bp,178.5bp) [draw] {strict};
  \node (f*adjoint2) at (464.95bp,526.0bp) [draw] {f.adjoint2};
  \node (f*basic) at (564.95bp,122.0bp) [draw] {f.basic};
  \node (c*rezk) at (387.95bp,583.0bp) [draw] {c.rezk};
  \node (c*set) at (806.95bp,294.0bp) [draw] {c.set};
  \node (l*set) at (838.95bp,468.0bp) [draw] {l.set};
  \node (c*initial) at (544.95bp,352.0bp) [draw] {c.initial};
  \node (f*default) at (316.95bp,583.0bp) [draw] {f.default};
  \node (f*equivalence) at (463.95bp,468.0bp) [draw] {f.equivalence};
  \node (c*discrete) at (298.95bp,294.0bp) [draw] {c.discrete};
  \node (c*cone) at (521.95bp,236.0bp) [draw] {c.cone};
  \node (c*default) at (293.95bp,410.0bp) [draw] {c.default};
  \node (c*pushout) at (297.95bp,352.0bp) [draw] {c.pushout};
  \node (l*default) at (636.95bp,526.0bp) [draw] {l.default};
  \node (l*functor) at (636.95bp,468.0bp) [draw] {l.functor};
  \node (precategory) at (494.95bp,10.5bp) [draw] {precategory};
  \node (c*terminal) at (462.95bp,352.0bp) [draw] {c.terminal};
  \node (l*limits) at (616.95bp,352.0bp) [draw] {l.limits};
  \node (default) at (316.95bp,639.0bp) [draw] {default};
  \node (c*functor) at (581.95bp,294.0bp) [draw] {c.functor};
  \node (c*order) at (142.95bp,236.0bp) [draw] {c.order};
  \node (c*indiscrete) at (453.95bp,294.0bp) [draw] {c.indiscrete};
  \node (f*exponential_laws) at (352.95bp,526.0bp) [draw] {f.exp.\ laws};
  \node (iso) at (494.95bp,66.5bp) [draw] {iso};
  \node (nat_trans) at (408.95bp,178.5bp) [draw] {nat.\ trans};
  \node (f*adjoint) at (556.95bp,410.0bp) [draw] {f.adjoint};
  \node (c*sum) at (329.95bp,236.0bp) [draw] {c.sum};
  \node (c*comma) at (671.95bp,236.0bp) [draw] {c.comma};
  \draw [->] (c*initial) ..controls (516.95bp,334.15bp) and (494.62bp,319.92bp)  .. (c*indiscrete);
  \draw [->] (c*default) ..controls (426.11bp,402.61bp) and (741.12bp,383.39bp)  .. (781.95bp,363.0bp) .. controls (820.11bp,343.94bp) and (860.59bp,317.08bp)  .. (834.95bp,283.0bp) .. controls (819.91bp,263.01bp) and (753.05bp,248.85bp)  .. (c*comma);
  \draw [->] (c*fundamental_groupoid) ..controls (91.332bp,267.8bp) and (92.58bp,240.49bp)  .. (107.95bp,225.0bp) .. controls (167.86bp,164.63bp) and (434.94bp,134.15bp)  .. (f*basic);
  \draw [->] (f*exponential_laws) ..controls (372.07bp,500.42bp) and (391.59bp,475.7bp)  .. (410.95bp,457.0bp) .. controls (447.61bp,421.57bp) and (495.71bp,386.22bp)  .. (c*initial);
  \draw [->] (category) ..controls (494.95bp,149.03bp) and (494.95bp,110.14bp)  .. (iso);
  \draw [->] (l*default) ..controls (611.95bp,507.88bp) and (593.49bp,494.49bp)  .. (l*adjoint);
  \draw [->] (c*default) ..controls (244.6bp,381.52bp) and (163.21bp,334.55bp)  .. (c*fundamental_groupoid);
  \draw [->] (c*rezk) ..controls (377.39bp,565.8bp) and (370.82bp,555.1bp)  .. (f*exponential_laws);
  \draw [->] (c*default) ..controls (267.82bp,387.69bp) and (246.05bp,363.52bp)  .. (254.95bp,341.0bp) .. controls (259.67bp,329.06bp) and (269.01bp,318.38bp)  .. (c*discrete);
  \draw [->] (l*default) ..controls (636.95bp,508.34bp) and (636.95bp,497.75bp)  .. (l*functor);
  \draw [->] (c*sum) ..controls (373.23bp,220.92bp) and (428.57bp,201.63bp)  .. (category);
  \draw [->] (l*functor_preserve) ..controls (680.14bp,448.01bp) and (625.02bp,431.0bp)  .. (f*adjoint);
  \draw [->] (l*limits) ..controls (659.0bp,334.66bp) and (698.6bp,316.92bp)  .. (708.95bp,305.0bp) .. controls (735.08bp,274.92bp) and (740.41bp,226.47bp)  .. (c*finite_cats);
  \draw [->] (c*discrete) ..controls (291.77bp,269.19bp) and (287.17bp,242.77bp)  .. (298.95bp,225.0bp) .. controls (314.36bp,201.76bp) and (343.93bp,190.11bp)  .. (nat_trans);
  \draw [->] (iso) ..controls (494.95bp,49.783bp) and (494.95bp,40.154bp)  .. (precategory);
  \draw [->] (l*limits) ..controls (580.51bp,337.59bp) and (555.46bp,324.37bp)  .. (540.95bp,305.0bp) .. controls (529.79bp,290.1bp) and (525.19bp,269.01bp)  .. (c*cone);
  \draw [->] (c*default) ..controls (339.93bp,383.34bp) and (406.57bp,345.03bp)  .. (417.95bp,341.0bp) .. controls (505.14bp,310.08bp) and (532.52bp,324.48bp)  .. (622.95bp,305.0bp) .. controls (625.15bp,304.53bp) and (627.42bp,304.02bp)  .. (c*product);
  \draw [->] (c*default) ..controls (344.33bp,392.71bp) and (396.31bp,374.87bp)  .. (c*terminal);
  \draw [->] (c*set) ..controls (813.46bp,262.0bp) and (821.7bp,201.13bp)  .. (790.95bp,168.0bp) .. controls (765.49bp,140.57bp) and (656.69bp,128.67bp)  .. (f*basic);
  \draw [->] (c*default) ..controls (202.28bp,389.94bp) and (37.151bp,349.58bp)  .. (5.9506bp,305.0bp) .. controls (-21.776bp,265.38bp) and (58.759bp,247.25bp)  .. (c*order);
  \draw [->] (c*product) ..controls (621.09bp,263.7bp) and (553.22bp,217.86bp)  .. (category);
  \draw [->] (l*functor) ..controls (637.26bp,450.34bp) and (637.44bp,439.75bp)  .. (l*colimits);
  \draw [->] (c*fundamental_groupoid) ..controls (136.98bp,273.73bp) and (167.86bp,259.52bp)  .. (groupoid);
  \draw [->] (f*equivalence) ..controls (495.12bp,448.56bp) and (516.44bp,435.27bp)  .. (f*adjoint);
  \draw [->] (f*attributes) ..controls (617.74bp,160.24bp) and (599.89bp,147.31bp)  .. (f*basic);
  \draw [->] (l*limits) ..controls (631.84bp,334.38bp) and (641.75bp,322.65bp)  .. (c*product);
  \draw [->] (c*opposite) ..controls (435.36bp,217.0bp) and (455.95bp,203.69bp)  .. (category);
  \draw [->] (l*colimits) ..controls (631.46bp,392.08bp) and (627.46bp,381.02bp)  .. (l*limits);
  \draw [->] (l*functor_preserve) ..controls (743.05bp,449.6bp) and (742.05bp,439.97bp)  .. (f*yoneda);
  \draw [->] (f*exponential_laws) ..controls (314.97bp,498.04bp) and (272.94bp,462.46bp)  .. (253.95bp,421.0bp) .. controls (239.1bp,388.58bp) and (233.64bp,373.66bp)  .. (247.95bp,341.0bp) .. controls (253.67bp,327.95bp) and (265.0bp,316.92bp)  .. (c*discrete);
  \draw [->] (f*adjoint) ..controls (610.71bp,392.58bp) and (665.38bp,374.87bp)  .. (f*examples);
  \draw [->] (c*pushout) ..controls (278.74bp,332.16bp) and (266.26bp,318.31bp)  .. (256.95bp,305.0bp) .. controls (246.1bp,289.49bp) and (235.88bp,270.58bp)  .. (groupoid);
  \draw [->] (c*set) ..controls (757.19bp,285.07bp) and (686.62bp,270.45bp)  .. (630.95bp,247.0bp) .. controls (613.32bp,239.57bp) and (610.85bp,233.97bp)  .. (593.95bp,225.0bp) .. controls (570.6bp,212.61bp) and (543.37bp,199.94bp)  .. (category);
  \draw [->] (c*default) ..controls (361.47bp,396.67bp) and (441.1bp,380.25bp)  .. (507.95bp,363.0bp) .. controls (509.85bp,362.51bp) and (511.8bp,361.99bp)  .. (c*initial);
  \draw [->] (c*functor) ..controls (529.52bp,276.72bp) and (475.49bp,258.92bp)  .. (c*opposite);
  \draw [->] (default) ..controls (296.15bp,621.36bp) and (283.43bp,607.94bp)  .. (277.95bp,593.0bp) .. controls (257.33bp,536.8bp) and (275.55bp,464.64bp)  .. (c*default);
  \draw [->] (c*functor) ..controls (641.33bp,276.5bp) and (706.54bp,256.11bp)  .. (712.95bp,247.0bp) .. controls (728.38bp,225.08bp) and (698.26bp,204.18bp)  .. (f*attributes);
  \draw [->] (f*adjoint2) ..controls (464.63bp,507.6bp) and (464.47bp,497.97bp)  .. (f*equivalence);
  \draw [->] (c*discrete) ..controls (274.03bp,275.93bp) and (255.71bp,262.65bp)  .. (groupoid);
  \draw [->] (c*cone) ..controls (490.23bp,219.86bp) and (457.05bp,202.97bp)  .. (nat_trans);
  \draw [->] (l*default) ..controls (669.58bp,508.48bp) and (696.14bp,494.22bp)  .. (l*functor_preserve);
  \draw [->] (l*adjoint) ..controls (583.25bp,449.17bp) and (602.02bp,435.73bp)  .. (l*colimits);
  \draw [->] (c*default) ..controls (399.55bp,401.25bp) and (591.34bp,383.6bp)  .. (651.95bp,363.0bp) .. controls (670.43bp,356.72bp) and (672.41bp,349.55bp)  .. (689.95bp,341.0bp) .. controls (720.98bp,325.88bp) and (758.17bp,311.56bp)  .. (c*set);
  \draw [->] (strict) ..controls (564.95bp,161.66bp) and (564.95bp,151.44bp)  .. (f*basic);
  \draw [->] (l*adjoint) ..controls (556.95bp,449.6bp) and (556.95bp,439.97bp)  .. (f*adjoint);
  \draw [->] (l*limits) ..controls (606.17bp,334.13bp) and (599.18bp,322.54bp)  .. (c*functor);
  \draw [->] (default) ..controls (390.33bp,613.09bp) and (541.16bp,559.83bp)  .. (l*default);
  \draw [->] (c*opposite) ..controls (406.93bp,217.21bp) and (407.48bp,206.75bp)  .. (nat_trans);
  \draw [->] (c*default) ..controls (319.86bp,392.49bp) and (335.17bp,379.13bp)  .. (341.95bp,363.0bp) .. controls (357.28bp,326.54bp) and (345.38bp,279.15bp)  .. (c*sum);
  \draw [->] (default) ..controls (316.95bp,621.91bp) and (316.95bp,612.17bp)  .. (f*default);
  \draw [->] (f*yoneda) ..controls (738.0bp,391.6bp) and (737.5bp,381.97bp)  .. (f*examples);
  \draw [->] (c*cone) ..controls (514.59bp,220.33bp) and (509.01bp,208.44bp)  .. (category);
  \draw [->] (c*comma) ..controls (640.65bp,219.18bp) and (609.27bp,202.31bp)  .. (strict);
  \draw [->] (l*set) ..controls (831.92bp,429.76bp) and (817.28bp,350.19bp)  .. (c*set);
  \draw [->] (l*functor_preserve) ..controls (707.88bp,447.91bp) and (682.27bp,434.02bp)  .. (l*colimits);
  \draw [->] (c*sum) ..controls (352.17bp,219.83bp) and (373.14bp,204.57bp)  .. (nat_trans);
  \draw [->] (c*pushout) ..controls (378.91bp,335.47bp) and (487.47bp,313.29bp)  .. (c*functor);
  \draw [->] (f*basic) ..controls (542.53bp,104.22bp) and (525.23bp,90.504bp)  .. (iso);
  \draw [->] (l*default) ..controls (696.5bp,512.6bp) and (760.15bp,497.12bp)  .. (812.95bp,479.0bp) .. controls (813.82bp,478.7bp) and (814.71bp,478.39bp)  .. (l*set);
  \draw [->] (l*limits) ..controls (592.45bp,343.78bp) and (587.03bp,342.21bp)  .. (581.95bp,341.0bp) .. controls (497.08bp,320.72bp) and (395.82bp,306.18bp)  .. (c*discrete);
  \draw [->] (f*default) ..controls (328.0bp,565.51bp) and (334.72bp,554.87bp)  .. (f*exponential_laws);
  \draw [->] (c*terminal) ..controls (460.21bp,334.34bp) and (458.57bp,323.75bp)  .. (c*indiscrete);
  \draw [->] (c*default) ..controls (295.15bp,392.61bp) and (295.85bp,382.49bp)  .. (c*pushout);
  \draw [->] (f*exponential_laws) ..controls (351.12bp,491.26bp) and (351.74bp,434.95bp)  .. (377.95bp,399.0bp) .. controls (390.36bp,381.99bp) and (410.88bp,370.39bp)  .. (c*terminal);
  \draw [->] (f*examples) ..controls (686.06bp,333.21bp) and (641.65bp,316.48bp)  .. (c*functor);
  \draw [->] (nat_trans) ..controls (456.38bp,161.32bp) and (505.51bp,143.53bp)  .. (f*basic);
  \draw [->] (c*comma) ..controls (622.06bp,219.79bp) and (563.63bp,200.81bp)  .. (category);
  \draw [->] (c*indiscrete) ..controls (439.12bp,276.08bp) and (429.46bp,264.41bp)  .. (c*opposite);
  \draw [->] (f*examples) ..controls (713.08bp,333.05bp) and (698.18bp,320.7bp)  .. (c*product);
  \draw [->] (f*examples) ..controls (759.92bp,332.42bp) and (776.53bp,318.85bp)  .. (c*set);
  \draw [->] (groupoid) ..controls (282.04bp,197.26bp) and (424.38bp,109.84bp)  .. (iso);
  \draw [->] (l*set) ..controls (821.18bp,460.06bp) and (816.94bp,458.37bp)  .. (812.95bp,457.0bp) .. controls (765.84bp,440.79bp) and (710.05bp,426.65bp)  .. (l*colimits);
  \draw [->] (f*exponential_laws) ..controls (390.55bp,506.35bp) and (416.12bp,492.99bp)  .. (f*equivalence);
  \draw [->] (c*order) ..controls (167.43bp,227.66bp) and (172.85bp,226.13bp)  .. (177.95bp,225.0bp) .. controls (297.88bp,198.49bp) and (331.22bp,211.6bp)  .. (451.95bp,189.0bp) .. controls (454.78bp,188.47bp) and (457.71bp,187.87bp)  .. (category);
  \draw [->] (c*finite_cats) ..controls (683.99bp,160.21bp) and (628.16bp,142.29bp)  .. (f*basic);
  \draw [->] (c*product) ..controls (591.14bp,274.45bp) and (493.32bp,248.76bp)  .. (489.95bp,247.0bp) .. controls (464.1bp,233.49bp) and (439.23bp,210.48bp)  .. (nat_trans);
  \draw [->] (f*exponential_laws) ..controls (311.55bp,507.79bp) and (288.9bp,495.17bp)  .. (272.95bp,479.0bp) .. controls (252.13bp,457.88bp) and (248.52bp,449.39bp)  .. (239.95bp,421.0bp) .. controls (222.1bp,361.84bp) and (218.43bp,331.32bp)  .. (256.95bp,283.0bp) .. controls (269.87bp,266.8bp) and (289.96bp,254.44bp)  .. (c*sum);
\end{tikzpicture}
\caption{The category theory part of the library (c = constructions, f = functor, l = limits)}
\label{fig:category}

\end{sidewaysfigure}

\section{Path Algebra and Cubical Methods}
\label{sec:cubical}


The core innovation in homotopy type theory is its new interpretation of equality.
In contrast to proof irrelevant Martin-L\"of type theory, we need to be careful about
choosing well-behaved equality proofs in the library since we might need to prove
lemmas about these proof objects themselves.
We want to maintain brevity using tactics and equational rewriting while making sure
that the generated proofs do not become unwieldy.

After defining equality on a type "A" in the library's prelude as an inductive
type family over two objects of "A" which is generated by the reflexivity witness
"refl : Π(x : A), x = x", we can provide operations and proofs for the basic
\emph{higher groupoid structure} of these ``equality paths'':
Concatenation "p ⬝ q" and inversion "p⁻¹" of paths as well as proofs about
associativity and cancellation.
These are constructed using the dependent recursor of equality which we call \emph{path
 induction} and which, for each "a : A", provides a function "Π(b : A) (p : a = b), P b p"
given the reflexivity case "P a (refl a)".
Likewise, we can prove the functoriality of functions with respect to equality:
For a function "f : A → B" and "p : a = a'" we define "ap f p : f a = f a'" by
induction on "p".
Using an equality "p : a = a'" in a type "A" to compare elements of two fibers in
a type family "C" over "A", we define the \emph{transport} of an element "x : C a"
along "p" as "p ▸ x : C a'".

For higher paths and dependent paths, we follow what Dan Licata calls the ``cubical
approach''~\cite{LicataBrunerie2015}. The basic notion is that of \emph{pathover}, or a ``path over a path'',
which compares elements "x : C a " and "y : C a'" in different fibers of a
type family over some path "p : a = a'" in the base type.
We define the type of pathovers above a base point "a : A" and "x : C a" to be the
type family "pathover C x : Π{a' : A}, a = a' → C a' → Type" which is inductively generated
by
\begin{lstlisting}
idpo : x =[refl a] x
\end{lstlisting}
where "x =[p] x'" is notation for "pathover _ x p x'".
This definition allows us to define a version "apd f p : f a =[p] f a'" of
"ap" for dependent functions "f : Π(a : A), C a".
It is also used by Lean to express the dependent eliminators for higher inductive
types (c.f. Sect.~\ref{sec:hits}).
To work with pathovers we provide a variety of operations and lemmas, analogous to the higher
groupoid structure of paths. Pathovers correspond to equalities in a sigma type.

For higher paths in a type, we use squares and squareovers.
Just like paths were defined as an inductive type family indexed by their endpoints
we define the squares in a certain type "A" as the type family indexed by four corners
and four paths between those corners, which is generated by some identity
square with "refl" on all its sides.
Squares arise naturally when you need to prove a pathover in an equality type, which is often
required when proving equalities involving higher inductive types.

Squareovers are dependent squares over a square. It takes as arguments a square in the base type and
four pathovers over the sides of this square. These correspond to squares in a sigma type.  We also
have a library of cubes three-dimensional equalities. We could generalize these to cubeovers, though
we didn't need those yet.

\section{Higher Inductive Types}
\label{sec:hits}

One novel idea in homotopy type theory is the introduction of \emph{higher inductive types} or
HITs~\cite[Chapter 6]{hottbook}. Higher inductive types are a generalization of inductive types. With inductive types you can
specify which terms or points are freely added to that type. In contrast, when defining a HIT, you
can specify not only the points in that type, but also paths and higher paths. For example, the
circle "S¹" is a HIT with one point constructor and one path constructor:\footnote{Although we use syntax
  inspired by the Lean syntax for inductive types, this is not valid syntax in Lean.}
\begin{lstlisting}
HIT circle : Type :=
| base : S¹
| loop : base = base
\end{lstlisting}
This means that the circle is generated by one point and one path "loop : base = base". There will
be more loops in the circle, such as "refl base" and "loop ⬝ loop" and "loop⁻¹", which are all
different. Higher inductive types have elimination principles analogous to those of ordinary inductive types.

The most commonly used proof assistants which have HoTT support (such as Coq and Agda) do not
support HITs natively. Just adding HITs as constants is not
satisfactory, because then the computation rules are not judgmental equalities.
Instead, users of Coq use ``Dan Licata's
trick''~\cite{Licata2011trick}. 
The idea is that to define a higher inductive type, one first
defines a private inductive type inside a module with only the point constructors, and then adds
the path constructors as axioms. One then defines the desired induction principle using the
induction principle of the private inductive type and adds the computation rules of this induction
principle on paths as additional axioms. Then the user closes the module, and the result is that
only the data of the higher inductive type are accessible, while the induction principle of the
private inductive type is hidden. This ensures that the computation rules
are judgmentally true for point constructors (but not for path constructors), but a disadvantage
is that inside the module inconsistent axioms were assumed, and one needs to trust that the code in
these modules does not introduce an inconsistency in the system.
In Agda the rewriting feature is used so that users can extend the kernel with judgmental rewrite
rules, though there are no checks for any rewrite rule declared in this way.

In Lean we follow an approach similar to Agda rewriting feature, by building in judgmental rewrite ryles.
However, we only extend
the kernel with the rewrite rules for two ``trusted'' higher inductive types, namely the
"n"-truncation and the typal quotient (quotient for short). The quotient is parameterized by a type
"A" and a family of types "R : A → A → Type". So ``typal'' (the adjective of ``type'') means that we
quotient by a family of types and not a family of mere propositions.
The quotient is the following HIT:
\begin{lstlisting}
HIT quotient (A : Type) (R : A → A → Type) : Type :=
| i : A → quotient A R
| e : Π{x y : A}, R x y → i x = i y
\end{lstlisting}
For the "n"-truncation and the quotient, we add the type formation rule, point and path
constructors, and induction principle as constants/axioms.\footnote{For the \lstinline{n}-truncation we treat
  the fact that the new type is \lstinline{n}-truncated as a ``path-constructor.'' In
  \cite[Sect.~7.3]{hottbook} it is explained that the fact that a type is \lstinline{n}-truncated can be
  reduced to (recursive) path constructors.} Then we add the judgmental computation rules for the point
constructors to the Lean kernel; the Lean kernel is extensible in such a way that certain new
computation rules can be added to it. After that, we add the computation rules on paths as axioms.
As remarked in Subsect.~\ref{sec:consistency}, we know that the resulting type theory is consistent,
because "n"-truncations and typal quotients can be reduced to pushouts,
and type theory with univalent universes closed under pushouts is modeled by
\cite{Cohen-Coquand-Huber-Mortberg2016}.

Given these two HITs, we define all other HITs in
the Lean HoTT library using just these two. Some reductions are simple, for example
the homotopy pushout of "f : A → B" and "g : A → C" is the quotient on type "B + C" with the
edges "R" defined as an inductive family with constructor
"Π(a : A), R (inl (f a)) (inr (g a))".
Proving the usual induction principle for the pushout is then
trivial. Given the pushout, we have defined the other usual HITs: the suspension, circle, join,
smash, wedge, cofiber, mapping cylinder and spheres.
In particular, we define "circle" as "sphere 1", which is "susp (susp empty)". We can then
\emph{prove} the usual induction principle for this type, and it satisfies the computation rules on
the point constructors judgmentally

We can also define HITs with 2-path constructors using quotients.
This uses a method similar to the hubs-and-spokes method described in
\cite[Sect.~6.7]{hottbook}. From the elimination principle of the circle it follows that for any
path "p : x = x" in type "A" we can define a map "f : S¹ → A" with "ap f loop = p" by circle
induction. Then we can prove the equivalence
\begin{lstlisting}
  (p = refl x) ≃ Σ(x₀ : A), Π(z : S¹), f z = x₀
\end{lstlisting}
This equivalence informally states that filling in a loop is the same as adding a new point "x₀",
the \emph{hub}, and \emph{spokes} "f z = x₀" for every "z : S¹", similar to the spokes in a wheel.
This means that in a higher inductive type, we can replace a 2-path constructor "p = refl x" by a
new point constructor "x₀ : A" and a family of 1-path constructors "Π(z : S¹), f z = x₀".

However, this does not quite define 2-HITs in terms of the quotient, since this family of path
constructors refers to other path constructors (via the definition of "f"), which is not allowed in
quotients. For this reason, we construct 2-HITs using two nested quotients. We first define a
quotient with only the 1-paths and the hubs, and then use another quotient to add the spokes.

For a formal treatment of this, we need the following inductive family, which are the paths in a graph:
\begin{lstlisting}
inductive path {A} (R : A → A → Type) : A → A → Type :=
| of_rel  : Π{a a' : A}, R a a' → path R a a'
| of_path : Π{a a' : A}, a = a' → path R a a'
| symm  : Π{a a' : A}, path R a a' → path R a' a
| trans : Π{a a' a''}, path R a a' → path R a' a'' → path R a a''
\end{lstlisting}
A \emph{specification for a \emph(nonrecursive\emph) 2-HIT} consists of a type "A" and two families
"R : A → A → Type" and "Q : Π{a a' : A}, path R a a' → path R a a' → Type".
Using this, we define the 2-HIT "two_quotient A R Q" with constructors
\begin{lstlisting}
HIT two_quotient A R Q : Type :=
| i₀ : A → two_quotient A R Q
| i₁ : Π{a a' : A}, R a a' → i₀ a = i₀ a'
| i₂ : Π{a a' : A} {r r' : path R a a'}, Q r r' →
       extend i₁ r = extend i₁ r'
\end{lstlisting}
where "extend i₁ r" is the action of "i₁" on paths in "R", e.g.
"extend i₁ (trans r₁ r₂) := extend i₁ r₁ ⬝ extend i₁ r₂".
We first define a special case where the 2-path constructor has only
reflexivities on the right hand side.
We call this "simple_two_quotient A R Q'", where "Q'" has type
"Π(a : A), path R a a → Type" and where
\begin{lstlisting}
i₂' : Π{a} {r : path R a a'}, Q r → extend i₁ r = refl (i₀ a)
\end{lstlisting}
As mentioned before, we define "simple_two_quotient A R Q" in two steps. We first define a type "X" with only the 1-path constructors and the hubs:
\begin{lstlisting}
X := quotient A R + Σ(a : A) (r : path R a a), Q' r
\end{lstlisting}
We then define "simple_two_quotient A R Q' := quotient X R'" where
\begin{lstlisting}
inductive R' : X → X → Type :=
| mk : Π{a : A} (r : path R a a) (q : Q' r) (x : S¹),
       R' (f q x) (inr (a,q))
\end{lstlisting}
with "f q : S¹ → X" defined by induction so that "ap (f q) (loop) = extend (inl ∘ e) r" for "q : Q' r".

We now prove the expected (dependent) induction principle, (nondependent) recursion principle, and computation rules for this two-quotient. The only computation rule which we did not manage to prove is the computation rule of the induction principle on 2-paths. However, this rule is not necessary to determine the type up to equivalence.

We then define the general version, "two_quotient A R Q", to be equal to
"simple_two_quotient A R Q'" where:
\begin{lstlisting}
inductive Q' : Π{a : A}, path R a a → Type :=
| q₀ : Π{a a' : A} {r r' : path R a a'},
       Q r r' → Q' (trans r (symm r'))
\end{lstlisting}
We then show that "two_quotient A R Q" and "trunc n (two_quotient A R Q)" have the right elimination
principles and computation rules. It (perhaps surprisingly) requires quite some work to show that the correct computation rules of the truncated version follow from the untruncated version.

This allows us to define all nonrecursive HITs with point, 1-path and 2-path constructors.
For example, we define the torus "T² := two_quotient unit R Q". Here
"R ⋆ ⋆ = bool", which gives two path constructors "p" and "q" from the basepoint to itself. "Q" is generated by the constructor "q₀ : Q (trans [ff] [tt]) (trans [ff] [tt])" where "[b]" is notation for "of_rel b". This gives a path "p ⬝ q = q ⬝ p".
We also define the \emph{groupoid quotient}: For a groupoid "G" we define its quotient as
"trunc 1 (two_quotient G (@hom G) Q)" where:
\begin{lstlisting}
inductive Q :=
| q₀ : Π(a b c) (g : hom b c) (f : hom a b),
       Q (g ∘ f) (trans f g)
\end{lstlisting}
If "G" is just a group (considered as a groupoid with a single object), then the groupoid quotient
of "G" is exactly the Eilenberg-MacLane space "K G 1".

We have also defined the propositional truncation just using quotients in
Lean~\cite{vandoorn2016proptrunc}. An extension of this construction to "n"-truncations has been given on paper~\cite{rijke2017join}.
If we formalize this generalization in Lean, it is possible to remove
"n"-truncations as a primitive HIT in Lean.

\section{Conclusion}
\label{sec:conclusion}

We have described the HoTT library for the Lean proof assistant, which formalizes many results in
HoTT, including higher inductive types, synthetic homotopy theory and category theory. It has a
large library of pointed types, and uses cubical methods for reasoning about higher paths. In the
future, we hope to make a HoTT mode for Lean 3, possibly using a version of cubical type
theory~\cite{Cohen-Coquand-Huber-Mortberg2016,Angiuli-Harper-Wilson2017}.

\subsubsection{Acknowledgments}

We wish to thank the members of the HoTT group at Carnegie Mellon
University for many fruitful discussions and Lean hacking sessions,
and in particular Steve Awodey and Jeremy Avigad who have been very
supportive of our work.
Additionally, we deeply appreciate all the times Leonardo de~Moura
fixed an issue in the Lean kernel to accommodate our library.
Lastly, we want to thank all contributors to the HoTT library and
the Spectral repository, most notably Egbert Rijke and Mike Shulman.

The first and second authors gratefully acknowledge the support of the
Air Force Office of Scientific Research through MURI grant
FA9550-15-1-0053.  Any opinions, findings and conclusions or
recommendations expressed in this material are those of the authors
and do not necessarily reflect the views of the AFOSR.

\bibliographystyle{splncs03}
\bibliography{leanhott}
\end{document}